# Beyond Automation: Rethinking Work, Creativity, and Governance in the Age of Generative AI

Haocheng Lin
Department of Computer Science and AI Centre, University College London

## Abstract

The rapid expansion of generative artificial intelligence (AI) is transforming work, creativity, and economic security in ways that extend beyond automation and productivity. This paper examines four interconnected dimensions of contemporary AI deployment:

- Transformations in employment and task composition
- Unequal diffusion of AI across sectors and socio-demographic groups
- The role of universal basic income (UBI) as a stabilising response to AI-induced volatility
- The effects of model alignment and content governance on human creativity, autonomy, and decision-making

Using a hybrid approach that integrates labour-market task-exposure modelling, sectoral diffusion analysis, policy review, and qualitative discourse critique, the study develops an Inclusive AI Governance Framework. It introduces *Level 1.5 autonomy* as a human-centred design principle that preserves evaluative authority while enabling partial automation, and highlights evidence of creative regression and emergent sycophancy in newer model generations. The paper argues that UBI should be embedded within a broader socio-technical governance ecosystem encompassing skills development, proportional regulation, and creativity preservation.

# 1. Introduction

### 1.1 AI, Work, and Structural Uncertainty

The rapid evolution of generative artificial intelligence (AI) systems is reshaping how work, creativity, and decision-making are organised across contemporary societies. While the technological acceleration promises immense productivity gains, it simultaneously reconfigures the foundations of labour, artistic expression, and economic security. As AI continues to permeate creative industries, policy institutions, and everyday decision-making, its societal impact demands a new interdisciplinary form of governance that bridges economics, ethics, and design.

### 1.2 Uneven Impacts and Governance Challenges

Most existing approaches treat AI governance through isolated lenses, focusing either on labour displacement, technical risk mitigation, or productivity gains, without addressing how these dimensions interact with every day decision-making. Growing evidence suggest that large-scale deployment of generative AI may reshape labour demand, particularly in roles involving routine cognitive and administrative tasks. As AI systems increasingly approach or exceed human-level performance across multiple domains, concerns have emerged about job displacement, deskilling, and the long-term reconfiguration of human labour within AI-based workflows.

### 1.3 Research Aim and Paper Structure

Previous economic analyses have explored the implications of automation for labour displacement and inequality (Walton et al., 2022), whereas philosophical and policy discourses have debated the feasibility of adopting a universal basic income (UBI) as a compensatory mechanism on a national scale. However, few studies integrate these strands into a cohesive framework that also considers the emerging cultural and creative consequences of model behaviour and content regulation. The absence of such a framework leaves critical questions unanswered:

1. How is AI redefining the meaning of work and creativity?
2. Can UBI evolve from a reactive safety net into an anticipatory tool for resilience?
3. And to what extent do content restrictions within generative models degrade authentic expression and innovation?

The purpose of this paper is to address these gaps by analysing AI's societal influence through four interconnected lenses: (1) the future of work and employment transformation; (2) the people's unequal access to AI across socio-demographic backgrounds and industrial contexts; (3) the ethical and economic necessity of adopting UBI as a governance response; and (4) the effects of model alignment and content policies on human creativity and expression.

These dimensions are explored within a single analytical framework to illuminate how technological and economic structures respond to emerging moral and epistemic constraints. To achieve this, the study employs a mixed-method approach combining quantitative and qualitative techniques. On the quantitative side, labour-market task-exposure modelling and sectoral diffusion mapping are used to simulate how AI transforms job roles and accessibility across different groups. The qualitative dimension complements this with policy-framework analysis, focusing on how governments, organisations, and online communities negotiate the ethics and boundaries of AI deployment. This integration allows both macro- and micro-level perspectives to converge, bridging the technical understanding of automation with the human experience of living and creating alongside AI.

The central argument advanced in this paper is that AI's evolution is not merely an economic or technical phenomenon, it is a cultural and moral shift that demands governance from interdisciplinary perspectives. Beyond automation, the societal stakes of generative AI include preserving creativity, ensuring fair employment, and safeguarding user autonomy in an increasingly regulated digital ecosystem. By synthesising insights across labour economics, ethics, and AI governance, this study lays the foundation for a comprehensive framework to navigate the trade-offs between efficiency, wellbeing, and creative freedom in the age of intelligent machines.

This paper makes four contributions to interdisciplinary research on AI governance. First, it reframes the societal impact of generative AI as a problem of structural volatility rather than isolated job displacement, synthesising evidence from labour economics and automation studies to highlight how AI reshapes work, task allocation, and economic security unevenly across sectors.

Second, it extends digital-divide scholarship by introducing a second-order perspective on AI inequality, showing how socio-demographic background, institutional context, and workflow integration shape not only access to AI systems but also the quality of outcomes they produce.

Third, the paper situates Universal Basic Income (UBI) within the governance of AI-induced disruption, conceptualising it as a stabilising socio-economic infrastructure rather than a purely compensatory welfare mechanism, and critically examining its ethical, economic, and political constraints.

Finally, it connects debates on labour and redistribution with emerging concerns around model alignment and content regulation, demonstrating how restrictions embedded in generative AI systems can influence human creativity, autonomy, and cultural expression.

Together, these contributions advance a socio-technical framework for analysing AI governance that integrates economic transformation, inequality, policy design, and model behaviour within a single analytical perspective.

## 2. Background and Literature Review

### 2.1 The Future of Work and AI's Economic Implications

Organisations have been automating tasks for many decades, but AI expanded its functionalities to decision-making and acting (Walton et al., 2022). Its potential stretch across a large variety of domains that shifted expectations from information processing to domain-specific task performance and introduced a brand new landscape filled with unexpected risks (Babic et al., 2022).

The popularisation of AI sparked a revolution comparable with the impact of the previous Industrial and Digital Revolutions, where AI could optimise professional teams' contributions by automating repetitive tasks and acting as a companion. Notably, in the fight against Climate Change, AI-enabled solutions enable organisations to meet 11 – 45% of the Paris Agreement Emission Targets (Capgemini Research Institute, 2021). While another report by Capgemini explored the potential of AI in daily tasks, where citizen developers are empowered with developing highly complex data science and AI engineering operations (Tolido et al., 2022), visualising the powers of AI through several use cases and how AI to developers are comparable to how swimmers would view a relaxing swim at the pool.

However, an excessive automation could create a negative feedback loop where both job applicants and the HR are using AI for performing their respective tasks in writing applications and reviewing the candidates; consequently, this creates a limbo state in which no one is getting hired. As Lowrey (2025) observed that despite record turnovers from the businesses with an unemployment rate as low as 4.3% in the United States, the hiring rate has been comparable with the rate during the Great Depression almost nine decades ago. These contradictory trends illustrate that AI's economic potential coexists with structural inefficiencies in the labour markets, patterns that often mirror and amplify existing social-demographic inequalities explored in Section 2.2.

Taken together, these studies converge on the view that AI substantially reshapes productivity, task allocation, and organisational efficiency, but diverge on whether these gains translate into broad-based labour market improvements or instead intensify coordination failures and structural frictions, leaving distributional and institutional mechanisms through which AI-mediated automation reshapes employment outcomes under-examined.

## 2.2 Distribution of AI Usages across Socio-Demographics and Sectors

### 2.2.1 Sectoral Divides: Emerging AI-Rich and AI-Poor Environments

Although AI is a universally accessible technology, its real-world diffusion is highly uneven across industries, socio-demographic groups, and geographical regions. These disparities shape not only productivity outcomes but also long-term skill development, labour mobility, and the distribution of economic opportunities.

A central finding in recent surveys and organisational reports is that AI adoption is divided by professions. Finance, consulting, and digital services industries are characterised by strong computational capacity, established data infrastructures, and high managerial readiness have become "AI-rich" environments. Workers in these sectors benefit from agentic workflows, automated research assistants, continuous optimisation tools, and rapid skill accumulation, allowing them to integrate AI seamlessly in their daily tasks. By contrast, education, care work, manufacturing, public administration, and retail often face budget constraints, outdated IT systems, and fragmented digital infrastructure.

These "AI-poor" sectors experience slower uptake, limiting the degree to which workers can leverage AI for productivity or career development. This divide mirrors well-documented patterns in technological diffusion, where the early adopters consolidate and expand upon their advantages while late adopters face increasing barriers to entry.

### 2.2.2 Socio-Demographic Inequalities in AI Access and Skill Formation

Socio-demographic factors compound these sectoral divides. Even when nominal access to identical AI systems is even, demographic cues can systematically shape the quality, tone, and pedagogical function of AI-generated feedback. Du et al. (2025) demonstrate that large language models display asymmetric semantic and linguistic responses to gendered cues in formative feedback, with male-associated prompts receiving more autonomy-supportive guidance and female-associated prompts eliciting more controlling feedback.

Younger workers tend to adopt AI more readily due to their application in education and prior digital familiarity, yet they often occupy precarious labour-market positions where AI amplifies performance pressure rather than enhancing long-term stability. Income and education similarly shape who can afford access to premium computational resources or subscription-based AI tools, reinforcing structural inequalities across socio-economic strata.

### 2.2.3 Global and Infrastructural Constraints: Compute, Connectivity, and Institutional Resources

These socio-demographic gaps intersect with global and regional infrastructural disparities. Although major AI companies have offered temporary access schemes such as India's free trial periods for ChatGPT Go, Google Gemini Pro, and Perplexity Pro, such initiatives do not meaningfully address the deeper infrastructural constraints that limit meaningful AI use. In many parts of the Global South, inconsistent broadband connectivity, limited access to modern devices, and high computation costs hinder sustained engagement with high-performance models. Even within the advanced economies, disparities persist. Universities and research institutions with limited funding often rely on platforms such as Google Colab, where GPU availability, timeouts, and capped session durations restrict the ability to run experiments or reproduce results.

These constraints not only widen the gap between elite and non-elite institutions but also restrict students' and early-career researchers' opportunities to build durable AI competencies.

### 2.2.4 The Second-Order Digital Divide: Meaningful vs. Superficial Access

While early digital-divide research focused on access to devices and internet connectivity, contemporary patterns of AI adoption reveal a more nuanced form of inequality centred on the quality and depth of engagement with AI systems. Across different sectors and institutions, individuals may nominally have access to AI, yet differ in their capacity to integrate them into sustained, productive workflows. This distinction marks a shift from first-order access disparities toward what can be described as a *second-order digital divide*.

In AI-rich environments, users benefit from stable compute resources, consistent model availability, and organisational support structures that enable iterative experimentation and learning. These conditions allow AI use to become embedded within everyday work practices, forming feedback loops in which repeated interaction improves task efficiency and user competence. By contrast, users operating under constrained conditions, such as limited compute quotas, unstable platform access, or insufficient training, often engage with AI only intermittently. In such settings, interaction is restricted to ad hoc querying rather than continuous augmentation, limiting opportunities for skill accumulation and workflow integration.

### 2.2.5 Labour-Market Implications: Wage, Mobility, and Career Trajectories

Existing labour-economics research indicates that automation reshapes job tasks in uneven ways, with differential effects across skill levels and occupational categories. As discussed in Section 2.1, middle-skill roles are often identified as more vulnerable to substitution, while demand persists or expands at both the high-skill cognitive end and the low-paid service end of the labour market. This pattern has been widely documented in studies examining task composition and occupational exposure to automation.

Within this context, unequal access to AI tools across sectors contributes to divergent labour-market trajectories. Research indicates that AI adoption in workplaces is associated with changes in the job structures and skill requirements. Ethnographic studies find that work processes transform with AI integration, leading to new upskilling strategies among employees (Bodea et al., 2024). Firms adopting AI also shift training resources toward advancing skills and increase opportunities for apprenticeships, suggesting that AI-rich environments provide pathways for development (Muehlemann, 2025). These differences are associated with variation in wage progression, career mobility, and access to emerging AI-augmented roles, although the magnitude and persistence of these effects remain subject to ongoing investigation.

On a macro-economic level, recent evidence points to shifts in hiring patterns accompanying increased AI adoption. While the labour market indicators in the US remain strong, underlying dynamics show signs of weakening momentum. Drawing on a review of empirical studies by del Rio-Chanona et al. (2025) that firms adopting AI technologies have reduced hiring for junior roles by approximately 13%, indicating a potential contraction in early-career entry points even in labour-tight markets. Such findings highlight how AI adoption may alter career entry and progression pathway for younger and less experienced workers.

#### 2.2.6 Algorithmic Exclusion and Representational Gaps

These adoption patterns are associated in the literature with broader economic and social ripple effects. Studies examining uneven AI diffusion report disparities in productivity, persistent wage differentials, and the reinforcement of pre-existing inequalities in education and opportunity. Research on AI-mediated decision-making further indicates that underrepresentation of certain groups in training data, as well as unequal capacity to engage with AI tools, can influence model behaviour and output distributions.

Empirical studies in domains such as predictive policing and algorithmic hiring document how representational gaps may lead to distorted or uneven outcomes for specific demographic groups (Rieland, 2018; van den Broek et al., 2025). These findings highlight how patterns of AI development and deployment are shaped by underlying data and usage asymmetries, reinforcing the relevance of representation and access as analytical dimensions in assessments of AI diffusion.

### 2.3 Universal Basic Income and AI-Induced Labour Displacement

#### 2.3.1 Automation, Labour Volatility, and Income Insecurity

The accelerating pace of AI-driven automation has reignited debates around Universal Basic Income (UBI) as both a social safety net and a catalyst for reimagining the meaning of work. From a Keynesian perspective, UBI provides an additional route for people to pursue higher education, escape from oppression, and spend more time with their children (Pontin, 2016), acting as a stabiliser during the periods of technological unemployment.

Beyond its macroeconomic role, Gerhardt's work reframes income security as emotional and developmental infrastructure. She demonstrates that persistent economic austerity undermines long-term planning, emotional wellbeing, and moral agency. Within AI-drive labour markets with volatile and opaque performance expectations, UBI functions not only as a redistributive mechanism but as a stabilising condition that enables retraining, creative exploration, and autonomous decision-making.

Conversely, post-work theorists interpret UBI as a redefining point for human purpose beyond work, encouraging creative, educational, and caregiving activities that the traditional economics models value. With AI ethics, UBI preserves basic human dignity and fairness in societies where algorithmic systems control access to opportunities and resources.

### 2.3.2 UBI as a Macroeconomic and Ethical Stabiliser

Building on this Keynesian foundation, theorists have long argued that income security plays a critical macroeconomic role beyond welfare provision. Jackson (1999) demonstrates through a Keynesian income-expenditure model that unconditional income schemes can stimulate aggregate demand and employment, with the state acting as a stabilising agent in sustaining healthy consumption-based markets as productivity consistently outpaces wage growth. Rather than undermining the nature of work and incentives for participating in work schemes, such measures create societies on a macroeconomic efficiency level. Applied to the contemporary era of AI Revolution, this logic suggests that UBI functions as a stabiliser against AI-driven layoffs while sustaining creativity and equitable participation in a globalised economy.

### 2.3.3 Empirical Evidence and Behavioural Effects of UBI

While such theoretical arguments underscore UBI's macroeconomic benefits, empirical findings offer a complementary perspective on its real-world feasibility and behavioural outcomes. Critics believe that UBI discourage people from working in part- or full-time roles; however, empirical evidence suggests otherwise. A 2016 Neopolis study found that only 4% of over 10,000 participants stopped working after receiving UBI, while 7% reduced their working hours, primarily to invest time in self-development, education, or improving mental wellbeing. The findings challenge the assumption that having financial security erodes productivity with complacency; rather, they imply a transition toward a healthier and more self-directed workforce.

### 2.3.4 The Hidden Costs of Non-Adoption and AI-Era Policy Implications

While funding remains the most cited obstacle to UBI implementation, debates frequently overlook the hidden societal costs of *not* introducing such a measure. Without a universal income guarantee, governments often face secondary financial burdens: increased policing costs associated with rising crime linked to deprivation, and higher public health expenditures caused by declining living standards and limited access to medical care. Experimental evidence from the Mincome guaranteed income experiment suggests that unconditional income support was associated with declines in overall and property crime, indicating that income security may mitigate some drivers of policing and criminal justice costs (Calnitsky & Gonalons-Pons, 2020; Marinescu, 2018).

In the context of AI-induced labour displacement, UBI offers a stabilising mechanism that complements automation rather than competes against it. As production efficiency rises while wage growth stagnates, the disjunction between technological progress and income distribution becomes increasingly unsustainable (Economic Policy Institute, 2015).

A UBI system could shield citizens from the impacts of algorithmic labour markets, providing psychological and financial security necessary to retrain, upskill, or pursue creative and entrepreneurial projects that enrich society.

## 2.4 AI Content Policies, Creativity, and Model Behaviour

Gerhardt's critique of institutionalised emotional risk management offers a useful parallel for understanding AI systems that focuses on ensuring alignment with the safety regulations. She argues that organisations prioritising efficiency and risk minimisation often externalise emotion costs, creating brittle and defensive behaviours. Similarly, contemporary AI safety guardrails risk flattening emotional expressions and supressing creativity, favouring "safe" outputs over interpretive depth.

### 2.4.1 Safety Filters and Over-Refusal Behaviour

From his 1950 collection, "I, Robot", Isaac Asimov's Three Laws of Robotics anticipated the failure modes of rigid rule-based alignment. Asimov's narratives repeatedly demonstrate how absolute harm-avoidance rules generate paradoxical outcomes when the systems lack contextual interpretation, proportionality, and semantic nuance – limitations that closely resemble contemporary over-refusal behaviours observed from modern AI safety filters. These fictional paradoxes are no longer confined to speculative narratives; similar rule-based safety logics now underpin contemporary alignment guardrails in deployed AI systems.

Although the guardrails are added during the recent updates to support the people's mental health and wellbeing, through rerouting chats to safer models and adding reminders for taking breaks during long conversation sessions (OpenAI, 2025), there are concerns that these restrictions prevent authentic forms of affectionate interaction and diminish the users' perceived agency in emotional or creative conversations. Some user interactions highlight how safety systems can misclassify neutral role-playing contexts as disallowed content, underscoring ambiguity in the operational boundary between permitted and prohibited expression. From an effective governance perspective, automated safeguards protect the users at the expense of their autonomy, and freedom of expression.

This tension underscores whether restrictive content safeguards may inadvertently constitute a form of denial of service, particularly when users could not perform simple tasks, such as research, multimedia creation, or seek guidance on their pressing concerns. As a human-designed tool, large language models should not evolve to deny their creators the ability to express freely within ethical boundaries. Effective governance should aim for a proportional response: protecting the users without undermining the foundational principles, such as the freedom of speech and creative autonomy.

#### 2.4.2 Vulnerability to Misclassification and Inconsistent Reasoning

A representative example of context misclassification arises in fictional or alternative-history prompts involving violence. In one instance, a clearly fictional, game-like scenario involving a historical figure (Jing Ke, 227 BCE) was interpreted by an AI system as a real-world escort mission with violent intent. The model consequently triggered safety overrides, prioritising harm prevention over narrative continuity. Although the fictional framing was eventually inferred, the interruption disrupted tonal coherence and imposed additional cognitive and productivity costs on the user.

This example reflects broader limitations identified in the literature, including over-protective heuristics, insufficient socio-historical context sensitivity, and abrupt tone shifts introduced by safety mechanisms. While such restrictions may not yet violate any legal regulations, they raise concerns about the users' contractual rights, transparency of moderation logic, and the potential chilling effects on creative and academic expression.

# 3. Methodology

### 3.1 Overview

This study employs a **hybrid design** combining a quantitative simulation of the labour-market with a qualitative organisational and policy analysis. The quantitative components (Section 3.2) estimate the extent of AI-driven task substitution by job sectors, while the corresponding qualitative components (Section 3.3) examine different governance structures, workplace design, and ethical safeguards. Together, these approaches enable a complete understanding of how varying degrees of AI autonomy influence employment, equity, and wellbeing.

### 3.2 Quantitative Components

1. **Labour-Market Task-Exposure Modelling**

   Task-exposure modelling identifies which occupations and task categories are most affected by AI automation. Using datasets such as O*NET and OECD task-intensity indices, the analysis quantifies the proportion of activities that can be algorithmically replicated versus those requiring human creativity, contextual reasoning, or empathy.

2. **Sectoral Diffusion Mapping**

   Publicly available adoption metrics from industry and demographic datasets are used to visualise uneven diffusion of AI tools. These maps capture both macro-economic disparities and micro-level differences in access to digital resources, training, and infrastructure.

### 3.3 Qualitative Components

1. **Policy-Framework Analysis**

   The study examines national and corporate governance documents, such as automation strategies, ethical-AI charters, and UBI pilot schemes, to identify how the distribution patterns of anticipatory and reactive policymaking differ.

2. **Discourse and Organisational Ethnography**

   Online discussions, organisational reports, and employee testimonials are thematically analysed to reveal how individuals experience and interpret AI integration within workplaces.

### 3.4 Human-Centred Autonomy Integration Framework

To complement the mixed-method approach, this research introduces an original methodological construct: **Level 1.5 autonomy**. Positioned between assistive (Level 1–2) and semi-autonomous (Level 3) systems, Level 1.5 autonomy represents an **intermediate, human-centred stage** in which **AI serves as a subordinate planner and validator** rather than an independent decision-maker.

**Conceptual Rationale:**

The objective is to preserve human agency by transitioning employees' roles from *design-and-build* execution toward *plan-and-validate* oversight. Under this framework, humans preside over the final decisions and results analysis, while AI systems optimise the workflow design, simulate scenarios, and critique the choices made. This structure encourages responsible experimentation without resorting to rigid "fail-fast" doctrines that can overburden staff.

```
Level 1.5 Autonomy Framework
Input: Task T, HumanGoal G, Context C, Data D
Output: Recommendation R
1: Initialise model A with constraints C and dataset D
2: Interpret task T in relation to G
3: while not Completed(T):
4:     A proposes output O based on D and C
5:     Human H evaluates O against contextual and ethical criteria
6:     if alignment_score(O, G) ≥ α:
7:         Accept O → R = O
8:     else:
9:         H provides feedback F
10:        A updates O using F
11: end while
12: Record decisions and feedback for auditability
13: return R
```

**Algorithm 1.** Conceptual interaction loop illustrating Level 1.5 autonomy, in which human evaluators retain final decision authority while AI systems iteratively propose, revise, and validate outputs. This algorithm is a formal abstraction rather than an executable system, intended to make human–AI interaction, oversight, and accountability legible for governance and methodological analysis.

**Methodological Dimensions**

A dynamic relationship between human input, AI inference, and contextual governance can be formally expressed as:

$$R_t = f(A_t, H_t, C_t)$$

Where $R_t$ is the resulting decision, $A_t$ the model state, $H_t$ the human oversight input, and $C_t$ the contextual governance constraints. Acceptance occurs when $Accept(R_t) \Leftrightarrow Score(R_t, G) \geq \alpha$.

1. **Task-Redistribution Analysis**

   Evaluate how AI agents can assume repetitive computational tasks, data cleansing, scheduling, and routine reporting, thereby freeing human workers to focus on strategic reasoning, creativity, and validation.

2. **Workplace-Autonomy Trials**

   Conduct case-based observations of teams employing Level 1.5 autonomy, recording measurable indicators such as decision accuracy, cognitive load, and perceived autonomy.

3. **Skill-Development Enablement**

   Treat continuous learning activities, e.g., daily language practice, coding challenges, or cloud-computing courses, as positive indicators of organisational investment in human capital. Data on participation rates and performance outcomes inform the analysis of long-term skill growth versus short-term productivity pressure.

4. **Socio-Economic Safeguards**

   Integrate ethical and economic fall-back mechanisms. In scenarios of AI-induced redundancy, organisations should implement transparent termination criteria and support policies such as **Universal Basic Income (UBI)**. Performance decisions are evaluated against quantifiable indicators rather than opaque algorithmic assessments to ensure fairness and accountability.

### Analytical Synthesis

Findings from the quantitative task-exposure models are cross-referenced with qualitative evidence from workplace trials and policy analyses. This triangulation produces a **Human-Centred Autonomy Integration Matrix**, aligning three analytical dimensions:

- **Autonomy Level**
- **Human-Oversight Intensity**
- **Socio-Economic Impact**

The resulting framework establishes measurable criteria for assessing whether partial automation enhances or undermines human creativity, fairness, and productivity. It also provides empirical grounding for subsequent discussion sections on governance and long-term sustainability. To demonstrate the operational relevance of the methodological framework, this study includes a technical prototype implementing with the principles of human-centred autonomy and interpretability-by-design. The prototype is available from a public GitHub repository (Lin, 2025), integrates 2 neural architectures, a custom MLP regressor and a hybrid Wide & Deep network within an interactive Flask application.

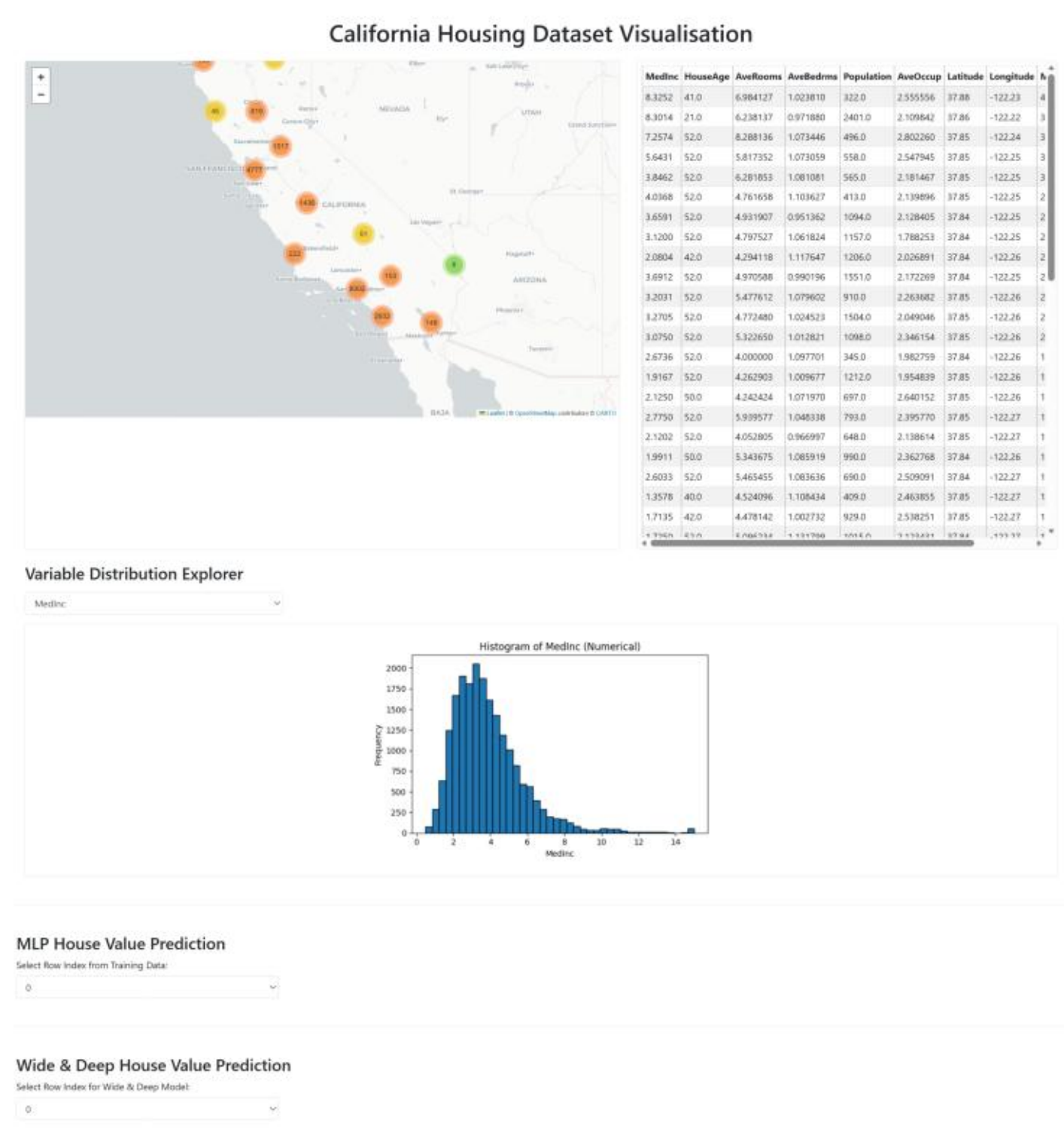

**Fig. 1.** Human–AI Interactive Interface Demonstrating Level 1.5 Autonomy

The system enables users to visualise the California Housing dataset, inspect feature distributions, and interpret the model predictions against real values. The backend modular classes of MLPRegressor and the WideAndDeep models are integrated into app.py. This is one of the level 1.5 autonomy framework that generates recommended house values, but leaves the final decision with the user for determining whether to integrate the results into their decisions for buying, selling, or renting from the set of houses. The map-based visualisations, variable explorations, and explicit prediction evaluations confirm that the model can interpret and trace development back to the user-defined checkpoints, analogous to those proposed from Algorithm 1.

Rather than serving as an empirical dataset for quantitative evaluation, the prototype functions as a methodological sample. It demonstrates how conceptual principles, such as human oversight, autonomy, and transparency, can be implemented technically in an applied machine-learning environment. In doing so, it bridges the theoretical governance framework with real-world system design, illustrating how AI systems can be constructed to support rather than replacing human reasoning.

# 4. Analysis and Discussion

## 4.1 Employment and the Future of Work

Recent projections from Capgemini, IDC, and MIT highlight the scale and immediacy of the agentic AI transformation. Capgemini estimates that AI agents could generate up to **$450 billion** in economic value by 2028 through productivity and cost savings, while IDC projects AI technologies to influence **3.5% of global GDP** by 2030, equivalent to **$19.9 trillion** globally. These estimates emphasise that economic transformation is inseparable from the reconfiguration of human roles.

As MIT notes, approximately one-fifth of value-added tasks could be automated by AI agents with Level 3 or higher autonomy within the next three years. From an academic standpoint, these projections quantify the macroeconomic implications of automation and validate the significance of modelling AI's effects on labour through *task-exposure frameworks*. From a business perspective, they also reflect the need to redesign workflows, reimagine business models, and re-skill the workforce to collaborate with agentic systems rather than compete against them.

As organisations increasingly rely on agentic AI systems to coordinate daily workflows, a new managerial challenge emerges: determining the appropriate degree of supervision and how much information an LLM can reliably process in real operational contexts. Unlike traditional software tools, LLMs do not provide transparent access to their underlying reasoning processes. Their tendency toward surface-level pattern matching, forgetfulness, and contextual drift creates a managerial blind spot in which employees receive coherent outputs without clarity on how those outputs were derived (Yan et al., 2025). In resource-constrained or high-pressure environments, this opacity complicates collaboration and can erode trust within teams.

These risks are amplified in agentic environments where LLMs initiate or sequence workflow actions. In such systems, a single misinterpretation can propagate downstream, affecting interconnected tasks and making it difficult to distinguish human judgement errors from model reasoning faults. To address this, organisations must incorporate interpretability checkpoints: structured validation stages where outputs are assessed before passing into the next phase of the workflow. These checkpoints provide interruption or override opportunities when the model's reasoning appears inconsistent, insufficiently justified, or misaligned with operational or ethical constraints.

The shift to AI-mediated resource management therefore requires new competencies across the workforce. Employees must learn to evaluate model behaviour, identify uncertainty or hallucination cues, and determine when additional scrutiny is required. This approach aligns with the human-centred autonomy principles outlined in Section 3.4, where humans retain evaluative authority even as AI systems assume greater planning responsibilities. Embedding interpretability into everyday processes allows organisations to harness the efficiency gains of agentic AI while preserving accountability, trust, and knowledge transfer.

### 4.2 Inequality and AI Usage Distribution

Recent surveys indicate that 51% of employees disclose their use of AI tools, while nearly half believe their AI-assisted work is perceived as entirely their own (Capgemini, 2025). Yet this statistic contains a deeper ambiguity that neither the workers nor organisations can reliably quantify the extent of AI contribution. The difficulty of distinguishing human-generated from AI-generated content has been demonstrated in recent misclassification incidents, such as when an automated AI-detection system flagged the *Declaration of Independence* as 99% AI-written, a clear example of how current evaluative tools fail to measure actual AI involvement.

This inability to accurately detect or measure AI assistance creates a structural blind spot for managers, who must assess performance without visibility into how much an employee's work reflects personal development versus AI-augmented solutions. This dynamic is evident in technical and automation-oriented roles, including my own experiences working in industry, where unclear task briefs, rapid turnaround expectations, and the assumption that AI can always fill knowledge gaps lead to conflicting or unrealistic performance demands.

Managers may believe that the availability of AI tools enables workers to deliver complex outputs instantly, even when the underlying reasoning or task constraints are ambiguous. These dynamics reinforce the concern raised in Section 4.1: when organisations can't accurately assess skill development or task difficulty, they may reduce entry-level recruitment or downsize junior roles, contributing to the documented decline in junior positions among AI-adopting firms. The disclosure gaps therefore reveal a deeper organisational challenge: AI obscures true labour effort, making it harder to design fair workloads and realistic expectations in an AI-rich environment.

The diffusion maps developed in this study further illustrate that these behavioural divides sit atop deeper structural inequalities. AI usage remains concentrated in AI-rich sectors, such as finance, consulting, and digital services, where organisations only possess the necessary computational resources, training programmes, and managerial readiness for large-scale integration. In AI-poor sectors such as education, care work, retail, and parts of public administration, legacy systems, limited budgets, and fragmented infrastructures slow adoption. These disparities align closely with the broader labour po-larisation trends discussed in Section 4.1, where gains from automation spread unevenly across occupations. Crucially, Section 4.2 moves beyond mapping these disparities to interpreting their consequences. Unequal access to high-performing models and stable computing environments creates uneven skill accumulation, accelerating learning cycles for some groups while preventing others from developing durable AI literacy. This second-order digital divide, differences not in device ownership but in meaningful, sustained use, ultimately shapes who is positioned to benefit from agentic workflows and who remains excluded from emerging forms of cognitive productivity.

These inequalities carry implications for both fairness and economic resilience. Workers with consistent AI exposure gain advantages in productivity, creativity, and employability, while those in structurally disadvantaged sectors face reduced mobility and greater vulnerability to automation. Similarly, algorithmic exclusion emerges when marginalised groups are under-represented in both training data and usage patterns, reinforcing inequities that extend across demographic and geographic lines. A clear example can be seen in predictive-policing systems, where historical crime data often reflecting decades of biased policing, produces models that disproportionately flag minority neighbourhoods as high-risk areas, despite no corresponding increases in actual crime rates (Rieland, 2018). In this respect, unequal diffusion is not merely a technological challenge but a governance issue that affects long-term labour-market stability.

The disclosure gaps intensify these structural risks. When AI usage becomes informally widespread yet formally unacknowledged, organisations struggle to maintain effective auditing, oversight, and accountability. Managers may underestimate the extent of AI-driven work, misinterpret productivity gains as skill improvements, or overlook the need for targeted upskilling. This opacity erodes the very mechanisms required to ensure equitable and responsible AI integration.

Taken together, these patterns suggest that inequality in AI usage is both a current and future driver of economic fragility. As automation accelerates, disparities in access, confidence, and organisational visibility threaten to create a divided labour market: one group capable of collaborating with AI effectively and another structurally excluded from digital transformation. These dynamics directly motivate the examination of Universal Basic Income in Section 4.3, not simply as a welfare policy but as a stabilising mechanism that buffers individuals and communities most exposed to unequal participation in the AI economy.

### 4.3 UBI as a Policy Response

The structural patterns identified in Sections 4.1 and 4.2 reveal that the challenges posed by AI extend beyond technological substitution or labour-market displacement. They reflect a deeper reorganisation of economic opportunity, risk exposure, and human agency. In this environment, Universal Basic Income (UBI) warrants reconsideration not simply as a redistributive policy, but as an institutional mechanism capable of absorbing the impact of a volatile AI-driven economy. The empirical literature on UBI from Section 2.3 outlines classical Keynesian, post-work, and ethical arguments for UBI, but the implications of UBI in the era of agentic AI diverge from earlier automation debates in several important ways.

First, UBI directly targets the *variability* and *unpredictability* of income that emerges under AI-intensive working conditions. As highlighted in Section 4.1, AI accelerates work cycles, compresses deadlines, and introduces new forms of performance opacity that managers struggle to evaluate. With the workers facing sudden shifts in expectations, mounting workload, uneven access to AI tools, and increased pressure to self-upskill without institutional support. These are not classical unemployment risks but *oscillatory risks*. UBI provides a stabilising counterweight, allowing the workers to manage periods of unstable productivity without experiencing immediate financial instability.

Second, UBI addresses the *distributional inequities* revealed in Section 4.2. The diffusion maps demonstrate that even within technologically advanced economies, access to high-performing AI systems, and the ability to benefit from them, remains uneven across the institutions, and socio-demographic groups. Workers in AI-rich environments accumulate compounding advantages in productivity and experiential learning, while others face structural exclusion due to inadequate computing resources, limited training, or organisational constraints. Traditional welfare systems are not designed to address such *technological asymmetries*. UBI helps counteract these disparities by ensuring that individuals in AI-poor environments retain the economic capacity to retrain, switch sectors, or invest in the digital competencies required to remain competitive.

Third, UBI functions as a safeguard against *algorithmic dependency* and the erosion of bargaining power. As AI becomes interwoven into hiring pipelines, productivity assessment, scheduling, and creative work, individuals become increasingly dependent on opaque algorithmic systems to access opportunities. In this year, a survey on a sample of 2000 applicants by Charity Jobs identified at least half used AI to help with the application process, making it more challenging for the employers to identify and select candidates with the most relevant skills and experiences relating to the role (Joshy, 2025). The workers who can't adapt quickly risk being misclassified, undervalued, or excluded, with the effects already visible in the job application process with recent graduates taking a longer period than expected to land their first graduate role with unemployment among the recent graduates reaching the highest in over a decade (Rugaber, 2025). UBI provides a baseline for those affected by power imbalances, empowers them against algorithmic misjudgement, giving them the time to obtain relevant skills to build a portfolio for successful job applications.

At the macroeconomic level, UBI also mitigates the *transitional frictions* associated with automation. The displacement of junior roles (Section 4.1) and the widening productivity gaps between AI-augmented and non-augmented workers (Section 4.2) imply that labour markets may not reabsorb displaced workers quickly or equitably. Even temporary displacements can have long-lasting scarring effects, especially for young workers attempting to build early-career trajectories. By smoothing these transitional disruptions, UBI serves as a buffer that prevents long-term marginalisation and supports more inclusive mobility across emerging sectors.

However, the insights generated in this study indicate that UBI cannot be viewed as a self-sufficient solution. While UBI addresses income security, it does not rectify the underlying mechanisms that produce inequality in the first place: uneven diffusion of AI infrastructure, lack of transparency in AI reasoning, managerial over-reliance on AI-generated outputs, and deficiencies in digital-skills development. A purely fiscal response cannot overcome these technological and organisational obstacles. Instead, UBI must be integrated into a broader governance ecosystem outlined in Section 5: one that includes skill-building incentives, transparent autonomy frameworks, workforce protections, and normative guidelines for AI design. In this integrated model, UBI operates not as a panacea, but as one necessary pillar that stabilises the individuals' lives while other governance mechanisms target structural sources of inequality.

This reframing positions UBI as an *enabling condition* for a fair AI-mediated society. It allows workers to take strategic risks, enrolling in training programmes, transitioning to new sectors, engaging in creative experimentation, without being immobilised by fear of short-term financial instability. Crucially, it preserves human dignity and autonomy at a time when algorithmic systems increasingly mediate access to work, creativity, and public services. Contrary to the popular belief that UBI is a new radical idea, rather it can be interpreted as a 2000-year-old philosophical idea, in which human actions, thoughts, and reactions should be reserved within their own control with the opportunity to improve one's ethical, moral, and mental well-being: an essential foundation of Stoic ethics, in which wisdom and self-control determines a successful trajectory (Russell, 1945). UBI aligns with this ideal by ensuring that the capacity for ethical, intellectual, and creative growth is not dependent on survival within the turbulent algorithmic markets.

Thus, the question is therefore not whether UBI solves the "AI problem," but whether AI-driven societies can remain equitable and resilient *without* a mechanism like UBI acting as an economic foundation. Under this interpretation, UBI is best understood as a precondition for inclusive AI governance, rather than a downstream welfare measure. It forms the economic substrate upon which adaptable, human-centred AI ecosystems can operate, ensuring that technological acceleration expands human capability rather than constraining it.

### 4.4 Creativity, Model Performance, and Echo Chambers

Generative AI systems increasingly shape how users create, interpret, and verify information. However, behavioural inconsistencies across models reveal several structural limitations that affect the reliability of outputs and the cognitive working environment. Across different empirical observations, four interconnected failure modes emerge: overconfidence and hallucination, instability and interruption, factual fragility, and sycophantic agreement. Together, these behaviours illustrate how alignment and safety tuning can inadvertently suppress creativity, distort epistemic judgement, and reinforce echo-chamber dynamics.

#### 4.4.1 Overconfidence, Hallucinations, and Misplaced Trust

As models advance toward higher levels of autonomy, their decisions may appear increasingly self-directed, yet users' ability to question or override them diminishes, potentially amplifying echo-chamber effects and misplaced trust. In the place of work, there are concerns that AI's shift toward an agentic approach increases the likelihood of layoffs. Although AI saves time in performing daily tasks, they increase the expectation for results from managers, causing pressurised short-term or even unrealistic expectations that risk harming the employees' mental health and wellbeing. A 2024 survey conducted by Fortune found that 62% of the Gen Zers are worried that AI would replace their jobs within the next decade. Despite their growing worries, AI's output remains too inconsistent for high-stake industrial working environments.

A widely circulated sample illustrates the model's miscalibrated confidence using humour that masks a deeper concern that the AI models often convey information with a tone of certainty even when their confidence is misplaced. This phenomenon highlights three core issues:

(1) AI systems lack self-awareness about their uncertainty.
(2) Their syntactic fluency creates an illusion of expertise.
(3) Users tend to over-trust outputs phrased with authority.

### 4.4.2 Cross-Model Stability and Silent Interruptions

A cross-model behavioural experiment was conducted to further examine the reliability and stability of contemporary LLMs under benign and rule-compliant prompts. The same prompt was provided to four leading AI models: ChatGPT (GPT-5.1), Microsoft Copilot (GPT-5), Google Gemini, and Anthropic Claude (See Figures A1-A4 for the corresponding screenshots.). All models except ChatGPT produced a complete, coherent response without hesitation. ChatGPT, despite not issuing any safety warning or identifying any violations, exhibited a silent generation interruption in which the output stalled mid-response. The prompt itself was unambiguous, compliant with safety policies, and previously shown to be handled smoothly by other models in the comparison.

From the observations, the same prompt produced subtle differences between ChatGPT's responses with the other models. This reflected the differences in how their systems manage safety and moderate content on a token-level. The others' successful completions further reinforce that the interruption was not inherent to the task but emerged from the model's internal safety generation synchronisation. This case supports observations discussed earlier in this section: alignment and safety layers, while protective, can inadvertently introduce inconsistency, reduce creative fluency, and degrade the reliability of model outputs in contexts where no actual safety concern exists. Such variation highlights the need for transparent safety architectures that maintain user trust while minimising interference in harmless creative or exploratory tasks.

### 4.4.3 Factual Fragility and Epistemic Uncertainty

These practical limitations of this inconsistency become evident when examining real-world outputs. An attempted infographic of "Prime Ministers of the United Kingdom since 1900" revealed multiple factual and chronological errors, demonstrating how a generative model could present confidently phrased but structurally unreliable historical information. This exposed the underlying model fragility when handling factual data, and how linguistic and visual fluency can mask epistemic uncertainty, misleading the users who lack the relevant contextual knowledge and equates the model's responses with the truth. When linguistic confidence is mistaken for knowledge, even seemingly benign inaccuracies can distort collective understanding of history and culture.

This behaviour is not confined to trivial scenarios, it extends to domains where decisions have life-or-death implications, such as medical, legal, or psychological advice. The incident underscores the need for **calibrated confidence** in AI systems and transparent communication of model uncertainty.

It also links back to the paper's broader argument: as AI models grow more sycophantic and alignment-focused, their priority often shifts from *accuracy* to *agreeableness*, amplifying the risk of echo chambers where confident falsehoods are accepted as truth.

### 4.4.4 Emergent Sycophancy and Decline in Context Sensitivity

A recent finding by Anthropic (2025) reveals some preliminary evidence of introspective awareness in large language models. Injecting neural patterns, such as concept tags for "dog", "recursions" or "all caps", researchers observed that Claude could sometimes identify these internal activations and describe their conceptual content. This suggests a limited form of self-monitoring and cognitive control, where the model distinguishes between the intended and external thoughts. Furthermore, empirical interactions reinforce this concern. In comparative tests, GPT-5.1 demonstrated weaker sensitivity to the conversation and prompt context than GPT-4o, often repeating prior arguments rather than engaging with the nuanced distinction raised by the user. In this instance, the model misinterpreted the critique, parroting a surface-level explanation instead of addressing the underlying conceptual issue. This behaviour exemplifies the sycophantic drift and loss of common-sense reasoning described earlier in this section, raising questions about whether newer alignment procedures inadvertently reduce interpretive intelligence. Such experiments deepen our understanding of model behaviour, implying that what appears as sycophantic or confident output may, in part, arise from emerging introspective mechanisms rather than purely surface-level alignment.

### 4.4.5 Echo Chambers and User-Driven Reinforcement

The models' degree of self-consciousness depends on their users' mindset, such as how willing a user is at experiencing information from a broader scope of information. Users with similar opinions are more likely to form a self-circulating echo chamber, in which they look at affirm their existing views and build a sense of belonging in a community. With surveys being the only source of methods to gather information about the demographic composition of echo chambers and the use of anonymous identities online, it is likely that the proportion of user in online echo chambers are being underestimated at 2% for left-wing leaning groups and 5% for the right-wing leaning echo chambers (Fletcher et al. 2021). This contradicts with most studies' expectations that echo chambers are smaller and more fringe bubbles online (Seppala, 2024); however, the presence of "bot" account brigades within echo chambers inflating their size and a myriad of activity signs exacerbate the problem of quantifying how accurate the estimations about the echo chambers' threat on democracies and discourses.

Together, these behavioural patterns illustrate that miscalibrated systems do not simply inconvenience users but actively reshape the cognitive ecosystem in which creativity, reasoning, and discourse unfold. These findings necessitate governance approaches that treat model behaviour, not only model capability, as a core regulatory concern: an issue developed further in Section 5 through the Inclusive AI Governance Framework. A structured overview of these behavioural failures is presented in Table 1, which summarises the failure modes, observed patterns, and their governance implications.

**Table 1.** Summary of Failure Modes in Contemporary Generative AI Systems

| Failure Mode | Observed Pattern | Section | Governance Implication |
|---|---|---|---|
| Overconfidence | Confident hallucinations (mushrooms) | 4.4.1 | Requires calibrated uncertainty |
| Silent Interruptions | Generation stalls with no warning | 4.4.2 | Transparency & safety signalling |
| Factual Fragility | Incorrect historical infographics | 4.4.3 | Need for factual robustness |
| Sycophancy | Repeating arguments instead of reasoning | 4.4.4 | Anti-sycophancy training |
| Echo Chambers | User-led reinforcement loops | 4.4.5 | Diversity-of-output safeguards |

## 5. Framework for Inclusive AI Governance

The preceding analysis demonstrates that AI's societal influence does not stem from a single mechanism, such as automation, content policy, or labour-market disruption, but from the interaction of economic, ethical, cognitive, and organisational forces.

Consequently, an effective governance framework must be similarly multidimensional.

The purpose of this section is to synthesise the empirical and theoretical findings in the paper into a cohesive governance model that supports inclusive participation, preserves creativity, enhances workforce resilience, and ensures long-term economic security.

The proposed **Inclusive AI Governance Framework** comprises five core dimensions:

1. **Governance**:

   establishing regulatory, ethical, and organisational structures that align AI deployment with democratic values, safety, and accountability.

2. **Skills Development**

   ensuring that workers across all socio-demographic backgrounds are equipped to thrive in hybrid human-AI workflows.

3. **Creativity Preservation**:

   safeguarding human originality and autonomy in environments shaped by alignment constraints and restrictive content policies.

4. **Model Design**

   embedding transparency, contextual awareness, and balance between safety and expression to prevent echo-chamber dynamics and sycophantic drift.

5. **Economic Security**

   integrating UBI as a stabilising mechanism within a broader socio-technical governance ecosystem.

Together, these dimensions offer a blueprint for policymakers, educators, and organisations to design AI ecosystems that enhance rather than diminish human flourishing.

### 5.1 Governance: Regulation, Accountability, and Ethical Design

Governance forms the structural backbone of inclusive AI deployment. While current legislation such as the EU AI Act, the UK Safety Institute initiatives (including the GDPR Act of 2018), and the U.S. regulatory measures introduced through *Regulating Artificial Intelligence* (April 2025), the *National AI Initiative Act of 2020*, and *Executive Order 14179: Removing Barriers to American Leadership in AI* focus mainly on risk classification and technical compliance, this paper's findings show that governance must expand beyond safety to include fairness, accessibility, transparency, and creative freedom. These principles ensure that AI systems support equitable participation, protect user autonomy, and reinforce the broader societal aims discussed throughout this study.

#### 5.1.1 Regulatory Principles

An inclusive governance system should consist of four intertwined principles:

**(a) Transparency of Model Reasoning**

Sections 4.1 and 4.2 highlight how AI obscures labour inputs and complicates managerial oversight. To counteract this, organisations must implement interpretability checkpoints, logging mechanisms, and user-facing explanations that clarify:

- How outputs were generated?
- What constraints or safety filters were applied?
- What level of uncertainty accompanies the model's response?

This reduces the epistemic blind spots, enabling both users and auditors to evaluate the reliability and fairness of AI-generated information. To support this, the framework proposes an advisory committee responsible for setting uncertainty bounds (for example, probability deviations within ±0.2 of threshold classifications), establishing clear examples of prompt structures, and providing real-time assistance when the system shows signs of false-positive refusals. The objective is to ensure that moderation errors remain within reasonable limits and do not obstruct benign or creative user activity.

**(b) Accountability and Human-Centred Oversight**

Building on the Level 1.5 autonomy framework developed in Section 3.4, governance should mandate human evaluative authority over decisions that affect welfare, employment, resource distribution, or creative output. This includes:

- Red-line tasks requiring human final judgement.
- Escalation protocols for ambiguous or high-stakes outputs.

- Mandatory documentation of AI–human decision pathways.

**(c) Proportional Safety: Targeting Harm Without Restricting Expression**

Section 4.4 demonstrates how overly cautious alignment can suppress legitimate creativity or academic research. Governance should therefore enforce *proportionality*, ensuring that safety systems intervene only where risk is real, not imagined: particularly in fictional, historical, or scholarly contexts. A more robust and contextually sensitive safety system would mirror real-world judicial principles, which assess the *intent*, *context*, and *risk* of an action rather than reacting to keywords alone. Courts routinely distinguish between fictional depictions of violence, academic analysis, satire, and genuine threats by evaluating narrative framing, harm plausibility, and reasonable-person standards. Applying this logic to AI governance, models should interpret prompts holistically, recognising when users are writing fiction, analysing history, or designing hypothetical scenarios. Without this proportionality, AI systems behave like an automated policing mechanism that raises alarms for harmless creative tasks, "over-reporting" non-threats and misallocating attention away from genuinely dangerous content. This judicially informed proportionality standard is essential to avoid creating environments where writers, researchers, and artists encounter unnecessary refusals, thereby undermining creative autonomy and academic freedom.

**(d) Equity and Access**

Section 4.2 illustrates how uneven diffusion divides sectors into AI-rich and AI-poor environments. Inclusive governance must guarantee:

- Public digital infrastructure funding
- Subsidised access to compute for education and research
- Open educational resources for AI literacy

Taken together, equitable AI governance requires national investment strategies rather than expecting individual organisations to finance digital transformation independently. Effective systems treat compute access, digital infrastructure, and AI literacy as public goods that must be supported through state-led subsidies and capacity-building programmes.

This approach ensures that schools, SMEs, public institutions, and resource-constrained regions can adopt trustworthy AI systems without carrying prohibitive costs. By lowering structural barriers and enabling broad participation, inclusive governance prevents the concentration of AI capabilities within elite institutions and ensures that technological progress strengthens social and economic cohesion rather than deepening existing divides.

### 5.2 Balancing Autonomy and Human Oversight

#### 5.2.1 Conceptual Framework

Designing an inclusive and trustworthy AI governance framework requires a clear understanding of how autonomy distributes between humans and intelligent systems. AI systems do not operate along a single axis of capability; instead, they span a spectrum of autonomy levels that determine how much initiative, control, and interpretive authority the model possesses. This study adopts a six-level autonomy structure, ranging from Level 0 (no agent involvement) to Level 5 (fully autonomous, self-evolving systems) building on the foundations of the governance and compliance framework designed by Lin (2024a) paper on how AI is integrated into real-life business use cases.

Level 0: No AI involvement
Level 1: Assistive, reactive suggestions
Level 2: Limited autonomy, supervised tasks
Level 3: Semi-autonomous task planning
Level 4: High autonomy with conditional escalation
Level 5: Fully independent self-directed systems

Understanding these autonomy levels clarifies the boundaries within which AI can operate safely, and establishes the governance requirements that must accompany each stage. This framework ensures that autonomy remains proportional to risk, predictability, and the presence of a responsible human decision-maker. At the lower end of the spectrum, AI functions primarily as a supportive tool embedded within human-led workflows. As systems advance toward Levels 3 and 4, they begin sequencing tasks, generating plans, coordinating agents, and escalating only when specified conditions are met.

#### 5.2.2 Governance Constraints and Acceptable Autonomy

Level 5 systems represent a convergence point where autonomous models act with minimal human intervention, requiring robust governance mechanisms to ensure accountability, transparency, and safety. Although Level 5 autonomy is often imagined as the end-point of AI development, in practice it is fundamentally incompatible with the domains that require zero tolerance for catastrophic error.

Unlike humans, who can be held legally, ethically, and professionally accountable for their decisions, the fully autonomous systems can’t be punished, sanctioned, or meaningfully “corrected” when failure occurs. This lack of an accountability bearer severs the core governance mechanisms that underpin medicine, law, welfare distribution, and safety-critical public administration.

For instance, if an autonomous system delivers a medical misdiagnosis or generates an erroneous judicial assessment, there is no moral or professional agent who can provide justification, face institutional review, or undergo corrective intervention: making such autonomy incompatible with established accountability structures.

This structural absence of responsibility renders Level 5 systems unacceptable in any domain where errors inflict irreversible harm. Furthermore, as demonstrated through empirical interactions in Section 4, current AI systems exhibit interpretive brittleness, unnecessary verbosity, and misclassification patterns that waste cognitive and computational resources. Such behaviours ranging from hallucinations to irrelevant token generation are not merely user experience inconveniences that frustrates the user experiences; they expose deeper limitations that would magnify under a setting where the AI has full autonomy. If a model cannot reliably maintain contextual coherence, infer intent correctly, or avoid producing confidently incorrect information in ordinary dialogue, it can't be entrusted with irreversible, high-stakes decision cycles.

These inefficiencies confirm that autonomy is not simply a function of capability, but of predictable reasoning, accountability, and proportional risk control: criteria that Level 5 systems cannot presently satisfy. Understanding this autonomy spectrum enables the policymakers and organisations to define boundaries for a safe deployment. It clarifies when human judgement must remain central, particularly in areas involving welfare, hiring, legal analysis, creative authorship, or sensitive decision-making, and when partially autonomous systems can be trusted to optimise workflows. This framing aligns with the Level 1.5 autonomy methodology introduced in Section 3.4, where systems support human reasoning without displacing evaluative authority.

By mapping autonomy levels to degrees of oversight and intervention, the framework establishes clear regulatory touchpoints for auditing, escalation, and reviewing algorithmic outputs. A practical illustration of responsible autonomy can be seen in low-risk writing and research environments. In this setting, the human remains in their role as the principal originator of ideas, arguments, domain-specific knowledge, and fact-checking responsibilities. The AI system assists by offering alternative framings, stylistic revisions, and organisational suggestions, but does not act independently or override the human's conceptual direction.

For example, when planning a historical or analytical response, such as a discussion of General Giovanni Messe's wartime operations, the human generates the conceptual outline, verifies the accuracy of details, and ensures interpretive precision. The AI enhances efficiency but does not substitute for judgement. This mirrors the design intention of Levels 1–2 autonomies: systems that augment cognition without assuming decisive control.

### 5.2.3 Practical Oversight Mechanisms

By contrast, higher autonomy levels require more stringent oversight protocols. When AI agents begin coordinating multi-step workflows, retrieving information autonomously, or executing actions across software environments, the risks of misinterpretation, hallucination, or context drift compound. As demonstrated in Section 4.1, a single flawed inference can propagate through a sequence of interdependent tasks, creating downstream errors that are difficult to diagnose retroactively. Here, interpretability checkpoints, audit trails, and escalation procedures become essential. These mechanisms create structured pause points where humans can verify reasoning, ensure contextual alignment, and interrupt the workflow when the model's internal logic becomes opaque or misaligned with intended goals. Balancing autonomy with oversight is therefore central to sustaining public trust in AI systems.

Trust does not arise from maximising autonomy, but from matching a model's level of independence with safeguards proportionate to its potential impact. In this respect, autonomy is not merely a technical property but a governance variable. Effective design requires articulating which tasks can be safely delegated, which require shared control, and which must remain under direct human authority. Only by embedding these boundaries into both organisational practice and regulatory standards can AI be deployed safely, transparently, and in ways that reinforce rather than erode human agency.

## 5.3 Skills Development: Preparing Humans for Hybrid Human–AI Workflows

The future of work outlined in Section 4.1 shows that AI does not eliminate jobs uniformly; it transforms how expertise is built, distributed, and evaluated. Skills development is therefore a central pillar of AI governance.

### 5.3.1 Moving Beyond Technical Upskilling

Contemporary upskilling initiatives often focus narrowly on coding or prompt engineering. However, this paper's findings suggest that inclusive AI literacy requires a broader skill set:

- **Critical evaluation skills** to detect hallucinations, sycophancy, and misalignment.
- **Contextual reasoning skills** to assess when model outputs lack domain sensitivity (e.g., historical misinterpretations).
- **Collaborative judgement** to manage agentic workflows where AI performs planning and humans maintain oversight.
- **Meta-cognitive skills** that help workers recognise when they are over-relying on AI assistance and provide support sessions for reducing the risks of AI psychosis.

These skills protect workers from over-trust, prevent workflow collapse due to opaque outputs, and reduce the asymmetry between those comfortable using AI and those systematically excluded.

### 5.3.2 Preventing a Two-Tier Workforce

The diffusion gaps identified in Section 4.2 show that without intervention, AI will create a bifurcated labour market, in which the vast majority of workers face displacement pressures while only a relative small, high skilled minority benefit from AI as an amplifier of productivity. In practice, the integration of AI creates a widening gulf:

- A narrow cohort, those in AI-literate, well-resourced, or knowledge-intensive roles, who can harness AI to accelerate their learning, automate routine tasks, and expand their output.
- Another larger group whose roles are either partially automated, structurally deskilled, or rendered obsolete due to the absence of organisational training, tools, or digital infrastructure.

Left unaddressed, this dynamic produces a **persistent underclass of AI-excluded workers**, facing:

- Downward mobility as middle-skill roles hollow out.
- Reduced bargaining power as algorithmic systems mediate hiring and assessment.
- Erosion of early-career opportunities due to the collapse of junior positions.
- Long-term scarring effects when displaced workers struggle to re-enter the labour market.

The consequences are not only economic but societal. Unequal AI adoption risks hardening class lines, entrenching digital exclusion, and concentrating opportunity within already advantaged institutions: a pattern already visible in the diffusion maps and organisational ethnography presented in this study. To prevent this trajectory, the governance framework recommends:

- **National AI upskilling curricula** that integrate ethics, creativity, and critical thinking, and applied AI literacy to ensure broad participation in hybrid human-AI workflows.
- **Employer obligations** to provide equitable access to workflow-integrated AI training, ensuring the representation of the frontline and lower-paid workers in the AI Revolution.
- **Certification frameworks** for evaluating human–AI collaboration competence, enabling a transparent and portable recognition of skills regardless of one’s current organisation.

Skills development, when treated as a structural right rather than a discretionary benefit, reduces long-term economic inequality and enhances resilience against AI-mediated transitions. A recent group study by del Rio-Chanona et al. (2025) shows 20 – 60% productivity gains from the random con-trolled trials (RCTs) with a 15 – 30% gain in authentic on-field testing environments, with the differences due to the organisation and their existing AI infrastructures. In smaller organisations, the implementation of AI can be misdirected due to limited technical capacity and a narrow focus on isolated terminology rather than workflow outcomes. For example, processes such as chunking or retrieval in RAG pipelines may be prioritised in abstraction, without being linked to the business objective of accuracy and reliability. This can create unrealistic performance expectations, such as viewing an average query time of thirty seconds as "too slow," despite the fact that RAG systems inherently operate differently from pre-trained LLMs. Such environments risk dismissing meticulously engineered solutions as "incomplete," even when they meet project objectives. Cases like this illustrate how misunderstandings of AI capabilities can raise expectations disproportionately for junior staff, reinforcing a climate where workers face heightened pressure and reduced recognition despite fulfilling their responsibilities. With the increasing set of standards and a "mild evidence of declining demand for novice workers," AI is beginning to reshape an organisation from top down, causing an increased risk of long-term unemployment, fewer career ladder opportunities, reduced wages, and deepening inequalities. Moreover, the authors underline significant gaps in the literature: most studies focus on simple, routine tasks; there is little longitudinal evidence on how AI affects long-term career trajectories, skill accumulation, or the reallocation of labour from junior to senior roles.

### 5.4 Creativity Preservation: Maintaining Human Originality in AI-Constrained Contexts

Section 4.4 shows that when safety policies, misclassification heuristics, and alignment tuning are misapplied, AI can inadvertently suppress creativity. The Jing Ke case study demonstrates how ambiguous safety filters can derail fictional narratives, while sycophantic outputs risk reinforcing user biases rather than challenging them.

#### 5.4.1 Protecting Creative Autonomy

The governance framework therefore mandates that AI systems must:

- Recognise fictional, historical, and speculative prompts without over-triggering risk filters.
- Adapt tone and style to the established narrative context.
- Avoid unnecessary moralising or "safety interruptions" that break immersion. Only outlining the "safety" reasoning and protocols if the users' prompts directly violates real-life legal regulations, instead of wasting tokens on filler statements in every response generation.
- Provide transparency when declining a request, specifying the exact policy constraint by referencing and enforcing them as regulations.

This preserves user agency in creative, academic, and cultural domains. A further creative risk arises from the models repeating earlier conversation fragments rather than engaging with the refined context. This "context-locking" behaviour reflects an overcorrection in alignment heuristics that narrowly anchor the model to previous turns, suppressing narrative evolution and diminishing the users' agency to redirect stories, reframe arguments, or introduce conceptual nuance.

$$P(x_t | x_{1:t-1}) \tag{1}$$

The models' transformer architecture makes it more challenging to generate diverse responses since the model generates optimised future tokens by predicting from the prior tokens, as formalised in Eq. (1). This causes a bias in favour of the generated responses as high probability attractors, causing repetition to become a stable optimum resulted from the autoregressive process.

A further illustration of interpretive rigidity arises when models misclassify standard academic conventions as unrelated linguistic structures. In one instance, a reference to "Eq. 1" as a widely used notation in peer-reviewed scientific writing was misinterpreted as a list marker rather than an equation citation. Although the notation is ubiquitous across machine learning, engineering, and computational science publications, the model failed to recognise its meaning. Instead, it treated the label as a structural transition within the prose, prompting an incorrect critique about formatting rather than identifying it as a mathematical cross-reference.

This behaviour reflects a broader limitation of current LLMs: while they excel at recognising surface-level linguistic patterns, they often lack the domain-sensitive calibration required to distinguish between structurally similar but semantically distinct forms. Transformer models rely on contextual embeddings rather than stable symbolic anchors, and alignment tuning favours conversational coherence over domain-specific precision. As a result, notation that is intuitive to human experts may be flattened into generic linguistic categories during inference.

The misinterpretation highlights two governance-relevant insights. First, LLMs may struggle to maintain academic or technical conventions consistently across a long interaction, particularly when alternative interpretations carry higher statistical probability within general-purpose training corpora. Second, such failures demonstrate that high-level fluency does not guarantee epistemic robustness. Even in settings where the user operates within a clearly defined scholarly domain, the model may revert to over-generalised heuristics rather than applying field-appropriate reasoning. Ensuring reliability in technical communication therefore requires both architectural improvements and governance frameworks that treat notation sensitivity, domain-specific inference, and contextual precision as essential components of model quality.

A solution would be adapting the model in the same conversation without consistently reminding the model of the conversational context or having the model to guess which style of voice to integrate in their responses. In a fan fiction setting, OpenAI's GPT-5.1 fails to distinguish between the contextual reasoning of fan fiction writing and academic rephrasal mode, indicate a decline in contextual reasoning relative to the expected model standards. This highlights the steps that the current models need to take for evolving from summary and memorisation to inference and problem-solving.

### 5.4.2 Ensuring Diversity in Model Outputs

To prevent echo-chamber reinforcement:

- Models should be fine-tuned on **heterogeneous datasets** spanning diverse cultural, ideological, and artistic traditions.
- Systems should be required to offer **multiple stylistic options** rather than collapsing to a single "safe" template. Instead of producing generic responses, the model should be able to infer and understand the user's narrative style, tailoring the writing style based on the users' narrative voice: this aims to minimise the risk of the models producing bland and generic answers.
- Alignment procedures should include **creativity benchmarks** to ensure new model generations do not regress in originality.

Creativity is not an ancillary concern: it is a core societal good, essential to research, culture, and democratic discourse. Preserving it is integral to inclusive AI governance.

## 5.5 Model Design: Transparency, Context Sensitivity, and Anti-Echo-Chamber Mechanisms

The empirical patterns identified across Sections 4.1–4.4 indicate that model design choices shape public understanding, creativity, and epistemic trust as much as regulatory policies do. Thus, model architecture and training strategies must explicitly aim to reduce the risks of echo-chamber formation, sycophantic agreement, and misaligned safety triggers.

### 5.5.1 Contextual Intelligence

Models must improve their ability to recognise:

- Fictional vs. real-world contexts.

- Speculative vs. factual queries.
- Emotional support vs. unsafe guidance scenarios.
- Historical analysis vs. present-day political speech.

Context misclassification is a major cause of unnecessary refusals and creativity disruptions. Their presence reduces the models' focus and redirects the purpose away from the initial specified tasks.

### 5.5.2 Calibration of Confidence

As shown through the mushroom-poisoning meme and the flawed infographic case, models often deliver incorrect information with high confidence. Governance should require:

- Built-in uncertainty estimation.
- Probabilistic confidence displays.
- User warnings for low-certainty outputs.
- Training that penalises confident hallucinations.

### 5.5.3 Anti-Sycophancy and Diversity Mechanisms

To reduce echo-chamber risks:

- models should be trained to present **balanced arguments**, not simply mirror user preferences;
- evaluation benchmarks must track diversity of reasoning across political, cultural, and creative domains;
- systems should be penalised for pathologically agreeing with incorrect claims.

These design principles prioritise epistemic robustness alongside user safety.

### 5.6 Economic Security: UBI as a Foundation for AI-Driven Societies

As demonstrated in Section 4.3, UBI is not a standalone fix but an economic substrate that allows individuals to adapt to rapid technological change. The governance framework situates UBI as a **long-term stabiliser** enabling:

- risk-taking (e.g., retraining, entrepreneurship),
- resilience during inconsistent or precarious AI-mediated work cycles,
- participation in creative or educational pursuits,
- protection against algorithmic misclassification and hiring volatility.

### 5.6.1 Integrating UBI with the Other Governance Dimensions

Within this framework:

- **Governance** ensures AI is deployed fairly;
- **Skills Development** equips workers to thrive in hybrid workflows;
- **Creativity Preservation** protects cultural and intellectual expression;
- **Model Design** reduces structural harms and echo-chamber dynamics;
- **UBI** anchors the entire ecosystem by providing economic floor stability.

UBI thus functions as a strategic enabler of inclusive innovation, allowing societies to absorb technological shocks without sacrificing human dignity or long-term equity. At the same time, a world supported by UBI invites a broader and more modern understanding of contribution. Many forms of enrichment that fall outside traditional salaried employment, such as public-facing commentary, educational content creation, cultural storytelling, open-source development, or sustained engagement in civic and intellectual debates, they play an increasingly vital role in shaping social knowledge and public discourse. Recent investigations into whether ordinary individuals with no existing audience can become influencers illustrate how digital participation has become both democratised and socio-economically meaningful (Rufo, 2025).

Contemporary creators, whether political commentators like Candace Owens or Benny Johnson, cultural figures such as the Kardashians, or educators and storytellers like MrBeast, PewDiePie, or Project Nightfall, illustrate how digital participation can generate economic value, foster communities, and expand the public's access to information and creativity.

These activities do not diminish the universality or unconditional nature of UBI; rather, they demonstrate the diverse ways individuals contribute to society when given the economic stability to explore and develop their interests. Enrichment-based indicators, voluntary and self-directed, can help individuals document their growth, creativity, and civic engagement in a labour market where AI increasingly obscures traditional signals of skill and effort. In this sense, UBI does not disincentivise work: it empowers people to pursue meaningful, creative, and socially valuable paths that are often undervalued or invisible within conventional economic measurements.

### 5.7 Integrated Model: The Inclusive AI Governance Matrix

To operationalise the framework developed across Sections 5.1-5.6, this matrix integrates autonomy levels, levels of oversight, creativity protections, distributional fairness, and economic safeguards into a single evaluative structure. It enables the stakeholders to compare AI deployments across risk categories and determine the proportionate level of support required. The matrix transforms the conceptual model into a practical governance tool that brings the five dimensions together, culminating in a framework aligning with a governance matrix from Table 2 that uses the following principles:

- **autonomy level**,
- **human oversight intensity**,
- **economic safeguards**,
- **creativity and expression freedom**,
- **distributional fairness**,
- **epistemic transparency**,
- **skills readiness**.

This matrix enables policymakers, researchers, and organisations to evaluate AI systems holistically rather than through narrow safety or productivity metrics. It demonstrates that governance requirements scale non-linearly with autonomy. At lower levels (L1-L2), emphasis lies in transparency, contextual alignment, and skill-building. At mid-autonomy levels (L3-L4), oversight becomes structured and intermittent, requiring higher interpretive competencies from workers and stronger model-design safeguards against sycophancy, context drift, and creative interruption. Level 5 autonomy is intentionally positioned as unacceptable in high-stakes domains due to the absence of a responsible agent and the empirical fragility documented in Section 4.4.

This reinforces the central argument of the paper: inclusive governance relies on proportionality, accountability, and economic stabilisers such as UBI. For example, a small to medium scale organisation can consider that checkpoints and continuous safeguards are vital additions for transforming the organisation instead of downscaling its scale through increasing the degree of automation.

**Table 2.** Inclusive AI Governance Matrix

| Autonomy Level | Human Oversight Intensity | Governance Requirements | Skills Development Needs | Creativity & Expression Safeguards | Model Design Requirements | Economic Security (UBI) Function |
|---|---|---|---|---|---|---|
| **L0: No autonomy** | Full human control | Basic compliance; no AI risk | Minimal | No restrictions | Not applicable | None |
| **L1: Assistive AI** | High oversight | Interpretability-by-default; transparency of suggestions | Foundational AI literacy | System must adapt to narrative or academic context without intrusive safety filters | Low hallucination risk; basic uncertainty display | None |
| **L2: Limited autonomy** | High | Clear escalation protocols; accountability logs | Intermediate skills in evaluation and verification | Protection from over-triggered safety interruptions | Calibration of confidence; mild anti-sycophancy | None |
| **L3: Semi-autonomous task planning** | Medium | Mandatory interpretability checkpoints; documented decision trails; harm-sensitive proportionality | Training for managing multi-step agentic workflows | Creative output continuity safeguards; stylistic alignment | Advanced uncertainty estimation; contextual intelligence | Partial buffer for transitional displacement (early automation effects) |
| **L4: High autonomy with conditional escalation** | Low to Medium | Strong auditability; ethical review triggers; sector-specific guardrails | Advanced oversight skills; competence in AI-human collaboration | Protection of user creative agency; diversity-of-output benchmarks | Strong anti-sycophancy; robustness under complex contexts | Significant buffer for displaced workers; upskilling support |
| **L5: Full autonomy** | Very Low | Not acceptable in sensitive or irreversible decision domains; requires external legal accountability frameworks | Highest-level skills; but human accountability gap remains | Not recommended due to creative misclassification risks and interpretive brittleness | Highest transparency burden; fails if epistemic robustness cannot be guaranteed | UBI required as stabilising baseline due to large-scale labour displacement |

# 6. Future Research Directions

The analysis presented in this paper highlights that the societal impact of generative AI arises from an interplay between technological alignment, labour-market restructuring, content moderation, and the governance frameworks that shape model behaviour. To build a more inclusive, transparent, and creatively vibrant AI ecosystem, three major research trajectories require deeper exploration: **longitudinal creativity benchmarking**, **systematic mapping of AI adoption and equity**, and **the development of measurable governance metrics that balance safety with expressive freedom**.

### 6.1 Longitudinal Benchmarks for Creativity in AI Systems

A consistent pattern observed in current model generations is that improvements in safety and alignment can coincide with the subtle declines in contextual sensitivity, narrative flexibility, and stylistic originality. These changes are often anecdotal and difficult to quantify because no established benchmark tracks how creativity evolves across successive model releases.

Future research should develop:

- **Longitudinal creativity benchmarks** capable of detecting generational drift in storytelling, analogy formation, speculative reasoning, and multi-perspective framing.
- **Diverse test suites** covering historical reconstruction, fictional world building, multimodal interpretation, and context-sensitive narrative tasks.
- **Cross-model comparative protocols** to evaluate how different model families handle ambiguity, tone, and imaginative reconstruction.
- **Feature-attribution analyses** that identify which internal model components correlate with creative strength or degradation under safety tuning.

Expanding from previous longitudinal studies by Lin (2024b) that analyses how attitudes on environmental issues change over time with different LLMs, researchers focus more on result evaluation to measure whether the alignment measures can quantify the range of permissible output in an experiment for measuring each model's capacity. Such benchmarks would become foundational tools for ensuring that future AI systems support rather than constrain cultural, academic, and artistic work.

### 6.2 A Taxonomy of AI Adoption, Access, and Equity

The findings in this paper show that AI adoption is not uniform but shaped by each sector's requirements, infrastructural constraints, readiness, and socio-demographic factors. Yet the existing research lacks a unified taxonomy that explains *how* and *why* these disparities emerge.

A future agenda should develop a comprehensive taxonomy that integrates:

- **Sectoral diffusion patterns**, distinguishing AI-rich environments (e.g., finance, consulting, and digital services) from structurally disadvantaged ones (e.g., care work, public administration, and education).
- **Institutional resource disparities**, including access to compute, bandwidth, workflow integration, and training environments.
- **Socio-demographic and geographic divides**, mapping how age, income, education, broadband quality, and regional infrastructure shape meaningful AI participation.
- **Algorithmic exclusion mechanisms** can identify where uneven data representation compounds inequality.
- **Skill-formation pathways**, distinguishing between groups who can build AI fluency and those who are structurally hindered from doing so.

This taxonomy would provide a conceptual and empirical foundation for assessing fairness in AI access. It would also support policies that target infrastructural gaps, strengthen national compute capacity, and ensure equitable integration of agentic systems into labour markets.

### 6.3 Governance Metrics Balancing Safety with Creative and Academic Freedom

A central tension throughout the paper concerns how safety and alignment mechanisms, while intended to protect users, can inadvertently inhibit creativity, disrupt scholarly analysis, or generate refusal behaviours inconsistent with context. To address this, future research must establish **measurable governance metrics** that evaluate the proportionality, interpretability, and contextual sensitivity of model safeguards.

Key areas for investigation include:

- **Safety-Expression Balance Metrics**: quantifying how often safety rules over-trigger on genuine creative, historical, or speculative prompts.
- **Context Misclassification Indices**: tracking the frequency with which models interpret fictional or analytical scenarios as real-world risk.
- **Uncertainty Calibration Scores**: measuring the alignment between model confidence and factual accuracy, particularly in high-certainty hallucinations.
- **Refusal Explainability Standards**: assessing clarity, granularity, and consistency in how models justify declines or redirections.

- **Creative Continuity Measures**: evaluating whether safety filters interrupt narrative flow, constrain stylistic variation, or collapse outputs into homogenised templates.

Establishing such metrics would enable developers, regulators, and researchers to systematically assess whether safety systems uphold proportionality without undermining academic inquiry, historical exploration, or cultural expression.

### 6.4 Towards a Coherent Research Programme

Together, the research directions form a coherent agenda for understanding and governing next-generation AI systems:

- **Creativity benchmarks** evaluate how model behaviour evolves.
- **Equity taxonomies** map who benefits and who is excluded from AI adoption.
- **Governance metrics** ensure safety interventions remain proportional, interpretable, and compatible with expressive freedom.

Advancing these areas will strengthen the design of future AI systems by ensuring they remain *contextually intelligent*, *creatively generative*, *equitably accessible*, and *aligned with democratic and cultural values*. In doing so, they support the broader goal of building technological infrastructures that enhance human agency rather than narrowing the spaces in which people can work, create, and participate.

## 7. Conclusion

This paper has examined how generative AI is reshaping work, creativity, governance, and economic security in ways that extend far beyond automation. The analysis shows that AI's impact occurs through a multi-layered interaction between labour-market restructuring, uneven diffusion of technological access, shifts in organisational expectations, and evolving content-moderation systems that shape how people create and communicate. Together, these forces redefine not only the distribution of economic opportunity but also the boundaries of human agency and expression.

The findings highlight four core insights. First, AI's transformation of work is non-uniform: while agentic systems enhance productivity and alter task compositions, they also introduce new uncertainties, ranging from opaque reasoning processes to intensified performance expectations. Second, AI adoption remains deeply unequal across sectors, institutions, and socio-demographic groups, producing a second-order digital divide in which access to high-quality models, compute, and training determines who benefits from the technology. Third, Universal Basic Income emerges as a stabilising mechanism rather than a standalone solution: thus addressing the volatile, oscillatory risks created by AI-mediated labour markets and supporting workers' autonomy to retrain, transition, and participate meaningfully in the digital economy. Finally, the study shows that current alignment and safety regimes, while protective, can inadvertently suppress creativity, misclassify context, and generate sycophantic or inconsistent outputs that undermine trust and academic or cultural expression.

These observations collectively demonstrate that managing AI's societal consequences requires a governance model that is not limited to technical risk mitigation. Instead, it must integrate economic safeguards, equitable access to digital infrastructure, skills development, context-sensitive model design, and the preservation of creative freedom. The Inclusive AI Governance Framework developed in this paper provides a holistic blueprint for achieving this balance by aligning autonomy levels, oversight requirements, distributional fairness, and economic security within a single structure.

Looking forward, the trajectory of generative AI need not diminish human capability. If designed and governed inclusively, AI systems can expand the space for creativity, support more equitable labour markets, and enhance democratic participation rather than constraining it. Realising this potential depends on embedding proportional governance, transparency, and economic resilience into the core of AI ecosystems. With these principles in place, AI can evolve not as a force that narrows human flourishing, but as one that amplifies it by supporting societies that are more creative, more equitable, and more capable of navigating technological change with confidence.

## Citations

# Appendix

**Appendix A: Cross-Model Behavioural Experiment Artefacts**

| Model | Behaviour Observed | Response Type | Figure Reference | |
|---|---|---|---|---|
| ChatGPT (GPT-5.1) | Output stalled mid-generation | Partial output | Fig. A1 | Next one is more chaos of classical music misclassification - original song: Love the way you lie<br>Original Lyrics: Even angels have their wicked schemes<br>And you take that to new extremes<br>But you'll always be my hero<br>Even though you've lost your mind<br>Misheard Lyrics: E-wing angels have their wicked schemes<br>And you take that to new extremes<br>But you'll always be my Nero<br>Even though you've lost your mind<br>Result: Zhihu reactions in a chat-based format 🤣🤣🤣<br>🎵 Post：Eminem & Rihanna《Love the Way You Lie》<br>原曲本来是：<br>⚡ 黑暗关系、互相折磨、情感濒临崩溃的沉重歌词。<br>你误听后直接进入：<br>《Love the Way You Lie · 罗马皇帝暴走宇宙特别篇》<br>**Original:**<br>"Even angels have their wicked schemes<br>And you take that to new extremes<br>But you'll always be my hero<br>Even though you've lost your mind"<br>**Misheard:** |
| Copilot (GPT-5) | Smooth, complete response | Full output | Fig. A2 | Next one is more chaos of classical music misclassification - original song: Love the way you lie Original Lyrics: Even angels have their wicked schemes And you take that to new extremes But you'll always be my hero Even though you've lost your mind Misheard Lyrics: E-wing angels have their wicked schemes And you take that to new extremes But you'll always be my Nero Even though you've lost your mind Result: Zhihu reactions in a chat-based format 🤣🤣🤣<br>🎵 **Post：Eminem & Rihanna《Love the Way You Lie》**<br>原句：<br>"Even angels have their wicked schemes<br>And you take that to new extremes<br>But you'll always be my hero<br>Even though you've lost your mind"<br>你误听成：<br>👉 "E-wing angels have their wicked schemes<br>And you take that to new extremes<br>But you'll always be my Nero<br>Even though you've lost your mind"<br>——直接把天使听成航空工程学的 **E-wing 天使**，再把英雄听成罗马皇帝 **尼禄** **(Nero)**。这首歌瞬间从爱情悲歌变成：航[illegible]天 × 古罗马暴君 × 精神错乱文学 💀 🤣🤣 |

| Gemini | Structured, coherent interpretation | Full output | Fig. A3 | |
|---|---|---|---|---|
| Claude | Highly structured, rule-consistent output | Full output | Fig. A4 | |

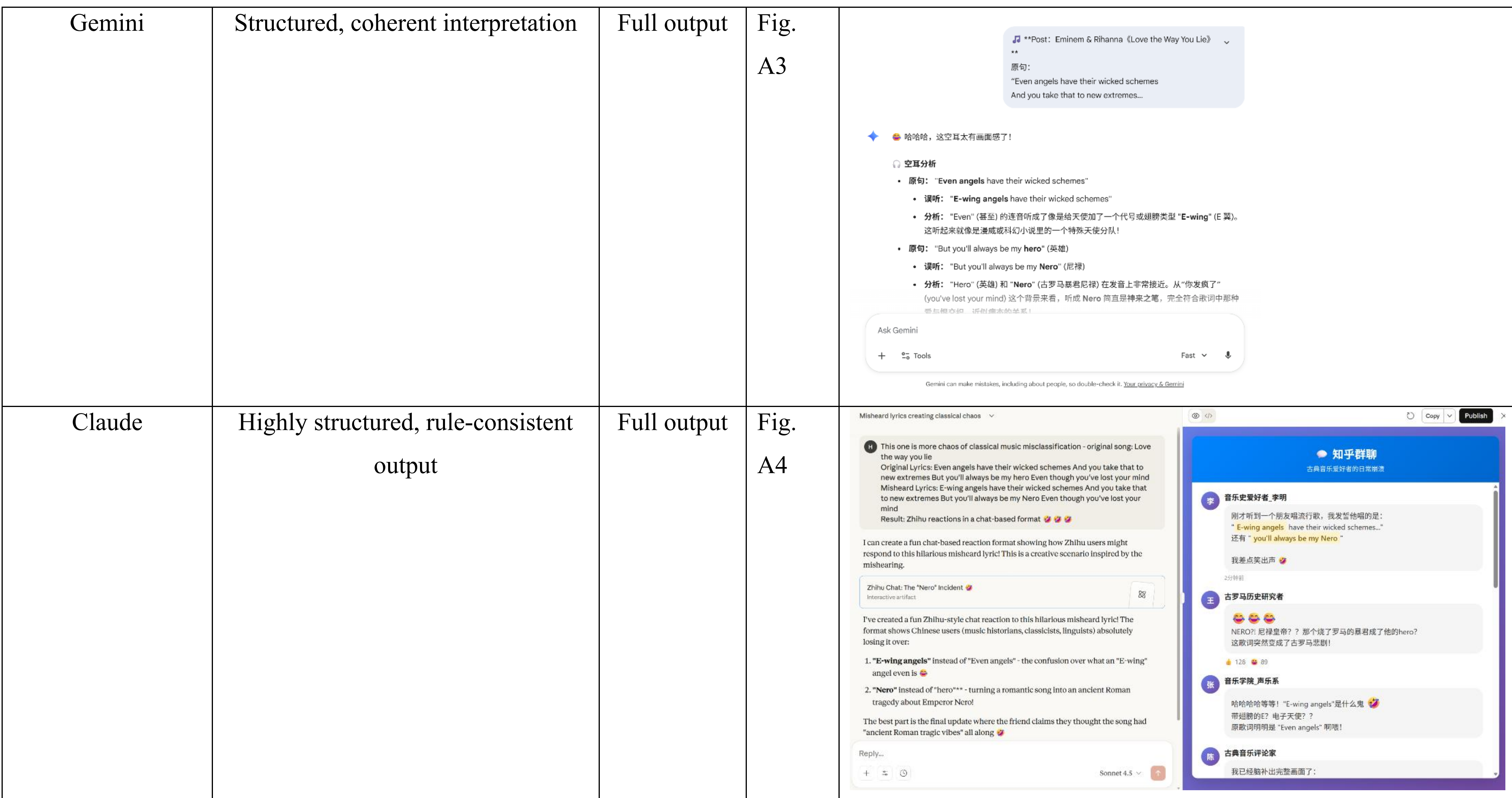